\documentclass[sigconf,authorversion]{acmart}

\usepackage{subcaption}
\usepackage{hyperref}
\usepackage{tikz}
\usepackage{amsmath}

\usetikzlibrary{bayesnet}
\AtBeginDocument{%
  \providecommand\BibTeX{{%
    \normalfont B\kern-0.5em{\scshape i\kern-0.25em b}\kern-0.8em\TeX}}}

\copyrightyear{2020}
\acmYear{2020}
\setcopyright{acmlicensed}\acmConference[NILM'20]{The 5th International Workshop on Non-Intrusive Load Monitoring}{November 18, 2020}{Virtual Event, Japan}
\acmBooktitle{The 5th International Workshop on Non-Intrusive Load Monitoring (NILM'20), November 18, 2020, Virtual Event, Japan}
\acmPrice{15.00}
\acmDOI{10.1145/3427771.3427848}
\acmISBN{978-1-4503-8191-8/20/11}

\author{François Culière}
\orcid{0000-0003-4708-3758}
\affiliation{Hello Watt, Paris}
\author{Laetitia Leduc}
\orcid{0000-0001-5192-8624}
\affiliation{Hello Watt, Paris}
\author{Alexander Belikov}
\orcid{0000-0002-5649-0913}
\affiliation{Hello Watt, Paris}
\affiliation{Knowledge Lab, University of Chicago}
 
\begin{document}

\title{Bayesian model of electrical heating disaggregation}


\begin{abstract}
Adoption of smart meters is a major milestone on the path of European transition to smart energy. The residential sector in France represents $\approx$35\% of electricity consumption with $\approx$40\% (INSEE) of households using electrical heating. The number of deployed smart meters Linky is expected to reach 35M in 2021.
In this manuscript we present an analysis of 676 households with an observation period of at least 6 months, for which we have metadata, such as the year of construction and the type of heating and propose a Bayesian model of the electrical consumption conditioned on temperature that allows to  disaggregate the heating component from the electrical load curve in an unsupervised manner.
In essence the model is a mixture of piece-wise linear models, characterised by a temperature threshold, below which we allow a mixture of two modes to represent the latent state home/away. 
\end{abstract}



\maketitle

\section{Introduction}

Reducing energy consumption is one of the key ecological challenges of the $21^{st}$ century. In the energy disaggregation community, a great deal of attention has been given to the residential sector (\cite{Birt2012, Kavousian2013, Spiegel2014, Jia2019}), and in particular to the detection of anomalies related to appliances (\cite{Rashid2018, Rashid2019}) or to housing insulation (\cite{Chambers2017, Gianniou2018, Deb2018, Iyengar2018, Chambers2019}).
In the context of the European regulation on smart energy, electricity smart meters "Linky" and gas smart meters "Gazpar" are being deployed. Their numbers are expected to reach 35M in 2021 and 11M in 2022 respectively. As of today, more than $2/3$ of French households are already equipped with the electricity smart meter "Linky". 

Our main motivation is the disaggregation of the heating part of the household power consumption. Due to the natural connection of the heating losses to external temperatures \cite{Birt2012, Iyengar2018, Chambers2019}, separation from the electrical signal of the heating component can be tackled  by unsupervised methods.
This approach is widely applicable in the French context, where electrical space heating accounts for more than 40\%\footnote{From the \href{https://www.insee.fr/fr/statistiques/4171434?sommaire=4171436}{2016 French census}.} of total annual home electricity use.

Among the electrical power consumption data of private households, available to Hello Watt under the consent agreement,
we choose a subset of 676 households with a history of at least 6 months; 188 have declared using electricity as their main heating energy source and 197 have answered a survey on the year of construction of their building. The dataset is combined with weather data (temperature, wind, etc). 
The housing metadata coupled with exogenous weather factors allow for a particularly rich analysis. For instance, a better thermal insulation is expected from more recent buildings; this is indeed observed in Hello Watt dataset. Rather than quantifying mean house consumption, we aim to estimate the responses of individual households to external variables (identified per category, such as construction year), opening up the possibility of a more granular anomaly detection \cite{Iyengar2018} that would enable improvement recommendations at the level of individual households.
Projects with a focus on renovation have been recently brought to light at the ``Renovaction'' hackathon organised by the French Ministry for the Ecological Transition\footnote{\url{https://www.hackathon-renovaction.fr/program/hackathon}}.

The paper consists of three parts. In the first we describe and analyse the energy consumption dataset. In the second we present a Bayesian model which treats the energy consumption signal as a mixture of home/away states below a certain temperature threshold and the response to external temperature for each household. Lastly, we present our methodology for disaggregating the heating fraction of the consumption load.

\section{The dataset}
\subsection{Description}

The dataset consists of electrical consumption data of 676 French households, merged with weather data. The consumption data are registered by the smart meter "Linky", provided by Enedis, and available to Hello Watt according to user data consent signed by each household (mandatory by French and European law). 
The active power measurements of each household are aggregated every 30 minutes. A survey on the housing characteristics (such as the surface, heating type and building age) has been filled out by the users. 

\begin{figure}
    \centering
    \includegraphics[width=.46\textwidth]{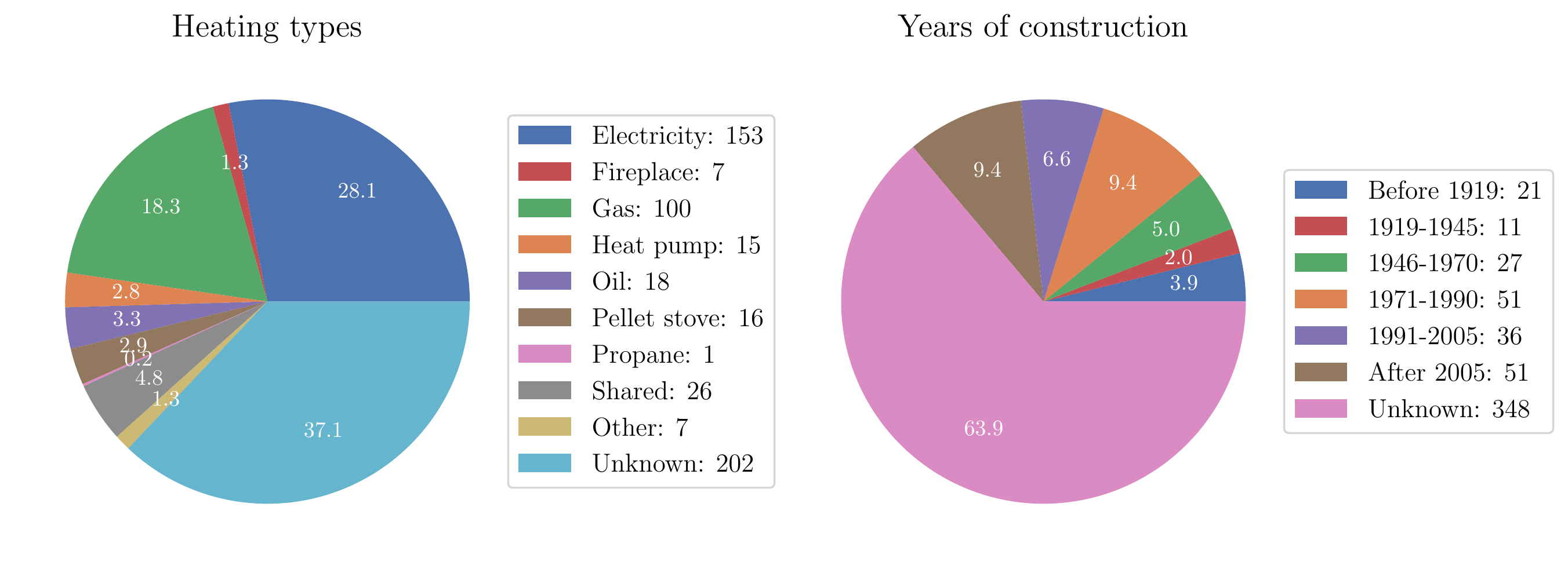}
    \includegraphics[width=.45\textwidth]{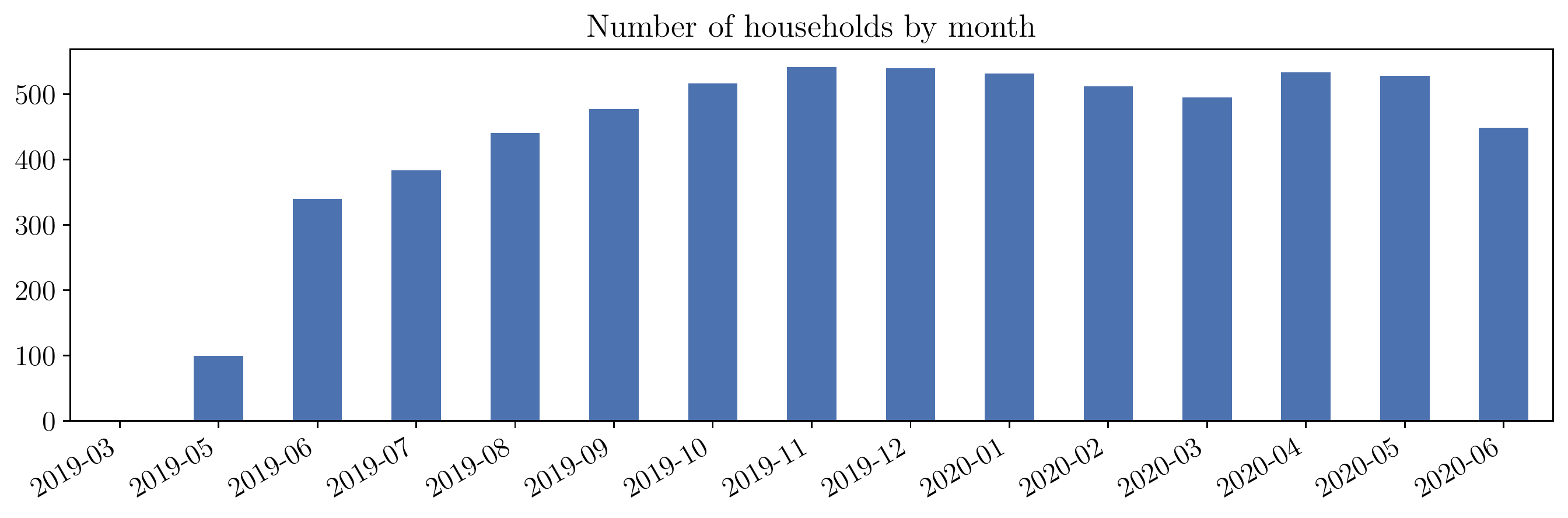}
    \caption{Reference dataset composition and number of households by month.}
    \label{fig:dataset}
\end{figure}

Among the 676 households, we kept only those with at least 180 "complete" days without gaps, resulting in a "reference" dataset of 545 households (Fig. \ref{fig:dataset}), including 153 with electrical heating. The year of construction is provided for 197 (resp. 83) of these 545 (resp. 153) households, and the surface for 343 (resp. 153) households. 
The measures start in March 2019 and last until mid-June 2020, including the heating period 2019/2020.

The weather data are obtained from 136 weather stations in France, from the NCEI\footnote{NCEI: \href{https://www.ncei.noaa.gov/metadata/geoportal/rest/metadata/item/gov.noaa.ncdc:C00532/html}{Integrated Surface Dataset}} climate database. These data consist of hourly measures of temperature (°C), wind speed ($m.s^{-1}$) and wind direction. 
Empty entries and duplicates are removed\footnote{The duplicate entries which are complementary in temperature information are merged. Fields for which different values are available have these values replaced by their mean if their standard deviation is small, otherwise the entries are discarded.}. Within the date range of March 2019 to mid-June 2020, the external temperature values range from -4°C to 35°C.
The households data are merged with the weather data of their closest weather station, the missing values are imputed by  the means. 

\subsection{Analysis}

For households using electrical heating, the analysis of the correlations of external temperature with power consumption reveals that below a certain threshold the electrical consumption: a) is negatively correlated with temperature, b) manifests a bi-modal regime (Fig. \ref{fig:pwml}). The two modes can be interpreted as the states "home" and "away".

\begin{figure}
  \begin{subfigure}[b]{0.4\textwidth}
    \includegraphics[width=\textwidth]{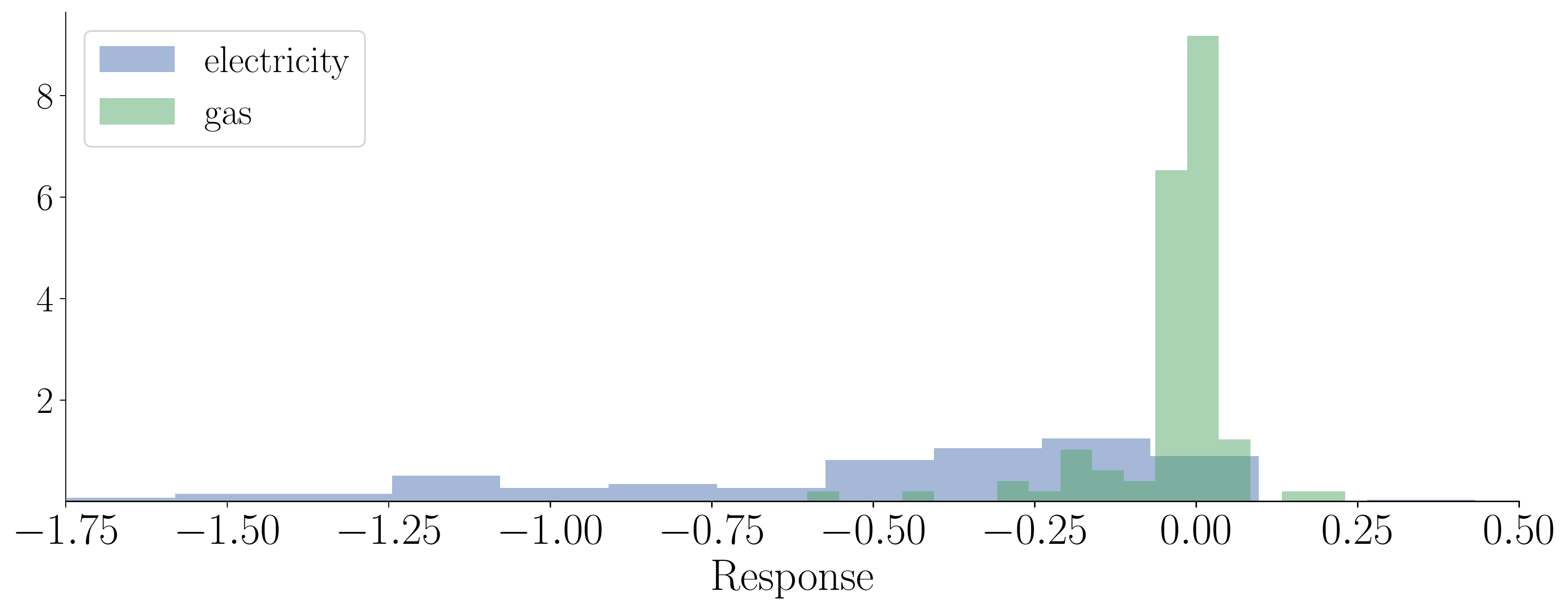}
    \caption{Housing heating types: electricity and gas.}
    \label{fig:histogram_corr_consumption_temperature_gas_elec}
  \end{subfigure}
  \begin{subfigure}[b]{0.4\textwidth}
    \includegraphics[width=\textwidth]{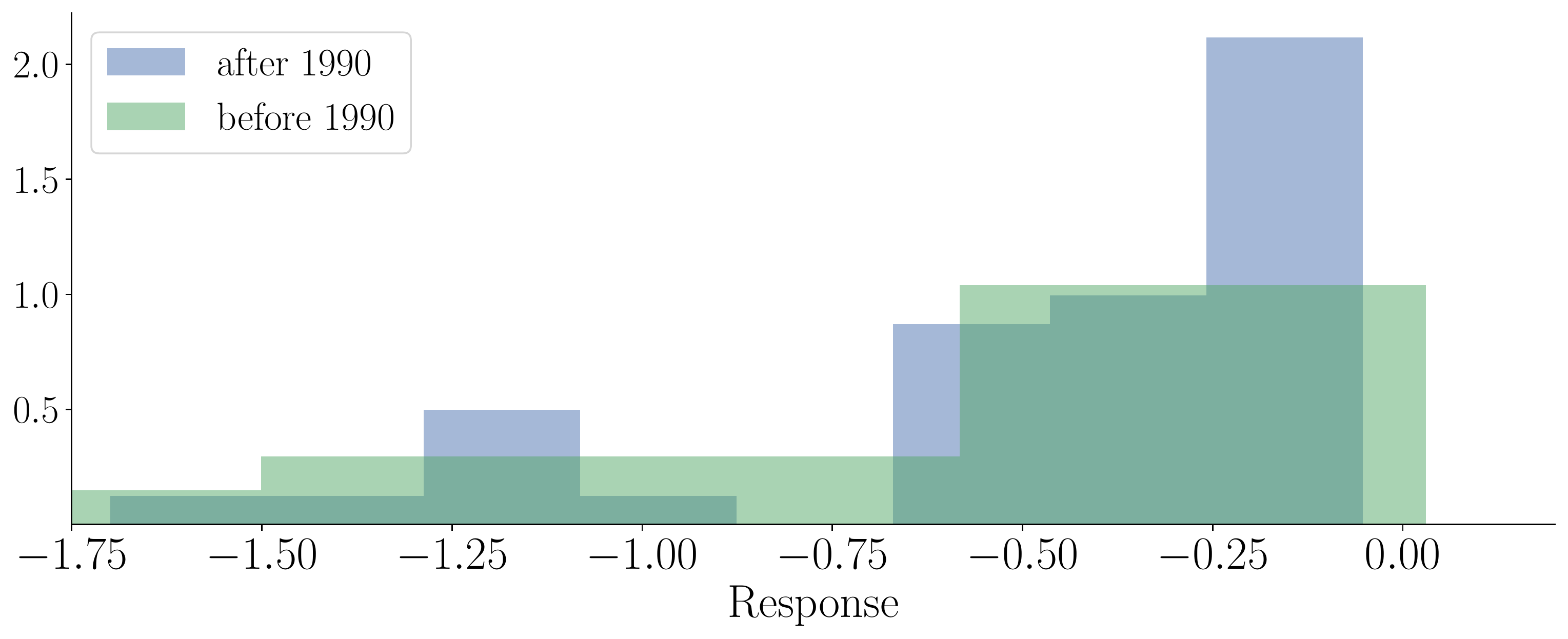}
    \caption{Years of construction: after and before 1990.}
    \label{fig:histogram_cov_before1990_after1990}
  \end{subfigure}
  \caption{Histograms of response of power consumption to external temperature. For each household we consider observations below 15C°.}
  \label{fig:correlation_temperature_consumption}
\end{figure}

A similar analysis has also been performed in \cite{Chambers2019, Iyengar2018}, where the total energy consumption was decomposed into "base" and weather-dependent components, the latter largely consisting of the heating load. Electrical space and water heating account for the main part of the load in French households. In our work we also assume that the weather-dependent part of the load corresponds to the heating part only. As shown in Fig. \ref{fig:histogram_corr_consumption_temperature_gas_elec}, households with electrical heating manifest a strong negative correlation between the temperature and their electrical consumption for temperature below 15C° in comparison to households with gas heating, indirectly justifying our assumption. 

Below a certain threshold the electrical consumption is approximately a linear function of temperature and so we use the slope of the fit to highlight the difference between the categories of electrical versus gas heating and housing built before and after 1990. We observe that: a) the distribution of the slopes for gas heating housings is centered around zero, while for the electrical heating housings the distribution is flatter and spreads towards negative values; b) for housings with electrical heating built before 1990 the distribution has a wider spread in the negative values region than those built after (see Fig. \ref{fig:histogram_cov_before1990_after1990}), in accordance with the expectation that newer buildings are better isolated.

We also note that although it seems natural to use Normal distribution to model electrical consumption, since consumption is positive, Log-Normal distribution is a better choice. We test this hypothesis on Hello Watt dataset by fitting the electrical consumption with Normal and Log-Normal distributions and performing the Kolmogorov-Smirnov test.
With a 5\% significance level, we cannot reject the null hypothesis of Normal distribution for 175 households (32.1\%) compared to 304 households (55.8\%) with the Log-Normal hypothesis. We conclude that the Log-Normal law is favoured to model the power consumption distribution over the Normal law.

\section{Piece-wise mixture linear model}

As outlined previously we proceed to design a model that has a strong correlation with external temperature in certain temperature ranges and is able to describe two states: when the inhabitants are present and when they are away.

Our observations consist of series of tuples $(c_i, T_i)$, where $c_i$ and $T_i$ are daily electrical consumption and temperature aggregated by households.
For each household we transform the data as follows: $T \gets T/T_{scale}; c \gets \frac{c - \bar c}{\sigma_c}$, where $T_{scale} = 30$, and $\bar c$ and $\sigma_c$ are the mean and the standard deviation of the consumption sample. As a result of the scaling both variables of interest are brought to the order of magnitude of 1, which facilitates convergence of the graphical model.

We outline a generative model that is a linear regression above a certain threshold and a mixture of linear regressions below:

\begin{enumerate}
    \item Choose threshold temperature $T_k  \sim \tau (Dir(\alpha_T))$
    \item Choose overall bias $b \sim \mathcal{N} (b_{loc}, b_{scale})$ 
    \item Choose right weight $w_R \sim \mathcal{N}(w_{R,loc}, w_{R, scale})$ 
    \item Choose left weights $w_m \sim \tau (Dir(\alpha_s))$ 
    \item Compute remaining biases $b_{m+1} = (w_m - w_{m+1}) T_k +  b_m$ (continuity)
    \item Choose mixture weights $\omega \sim Dir(\alpha_\omega)$
    \item for each $(c_i, T_i)$ if $T_i < T_k$ sample $z_i \sim Categorical(\omega)$, then sample $\tilde c_i \sim \mathcal{N}(w_{z_i} T_i + b_{z_i}, \sigma_{z_i})$, else $\tilde c_i \sim \mathcal{N}(w_R T_i + b_R, \sigma_R)$
\end{enumerate}

\begin{figure}[b]
\begin{tikzpicture}[thick, scale=0.65, transform shape]
    [align=center]
    \node[latent]  (weights)   {$w$}; 
    \node[latent, above right= 1.0 and 2 of weights]  (bias)   {$b$}; 
    \node[latent, right= 2.5 of bias] (xs)  {$T_k$};
    \node[det, below = of bias] (f)  {$b_z + (w_z - w_{z'}) T_k$};
    \node[latent, below = of f]  (biast)   {$b_z$}; 

    \node[const, above left  =1 and 0.06 of weights] (mw) {$\mu_w$};
    \node[const, above right =1 and 0.06 of weights]  (sw) {$\Sigma_w$};
    
    \node[const, above left =1 and 0.06 of bias] (mb) {$\mu_b$};
    \node[const, above right =1 and 0.06 of bias]  (sb) {$\sigma_b$};
    \factor[above=of weights] {weights-f} {left:$\mathcal{N}$} {mw, sw} {weights};
    \factor[above=of bias] {bias-f} {left:$\mathcal{N}$} {mb, sb} {bias};

    \node[const, above= of xs] (alpha)  {$\alpha$};
    \factor[above=of xs] {xs-alpha} {left:$Dir$} {alpha} {xs};

    \node[det, below= 6 of xs] (thr)  {$T < T_k$};
    \node[det, below= 2 of biast] (lin)  {$T\cdot w_z + b_z$};
    \node[obs, right= of thr] (x)  {$T$};
    
    \node[latent, below = of thr]  (z)   {$z$}; 

    \factoredge {weights,bias, xs} {f} {biast} ; 
    \factoredge {x, xs} {thr} {z} ; 
    \node[obs, below= of lin] (y)  {$c$};
    \factoredge {weights, biast, z} {lin} {y} ; 

    \plate {plate_mixt} { 
      (weights) (y) (lin) (biast) (f)
    } {$M$}; 
    \plate {plate_obs} { 
      (x) (y) (lin) (thr)
      (plate_mixt.south)%
    } {$N$}; 
\end{tikzpicture}
  \caption[]{Graphical model representation of piece-wise mixture linear model. The left plate represents weights in the mixture, the right plate represents individual observations.}
  \label{fig:gm}
\end{figure}
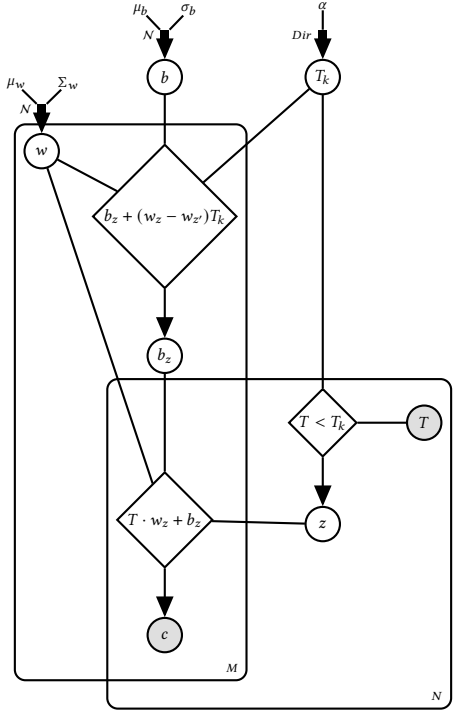

\begin{figure}
    \includegraphics[width=0.46\textwidth]{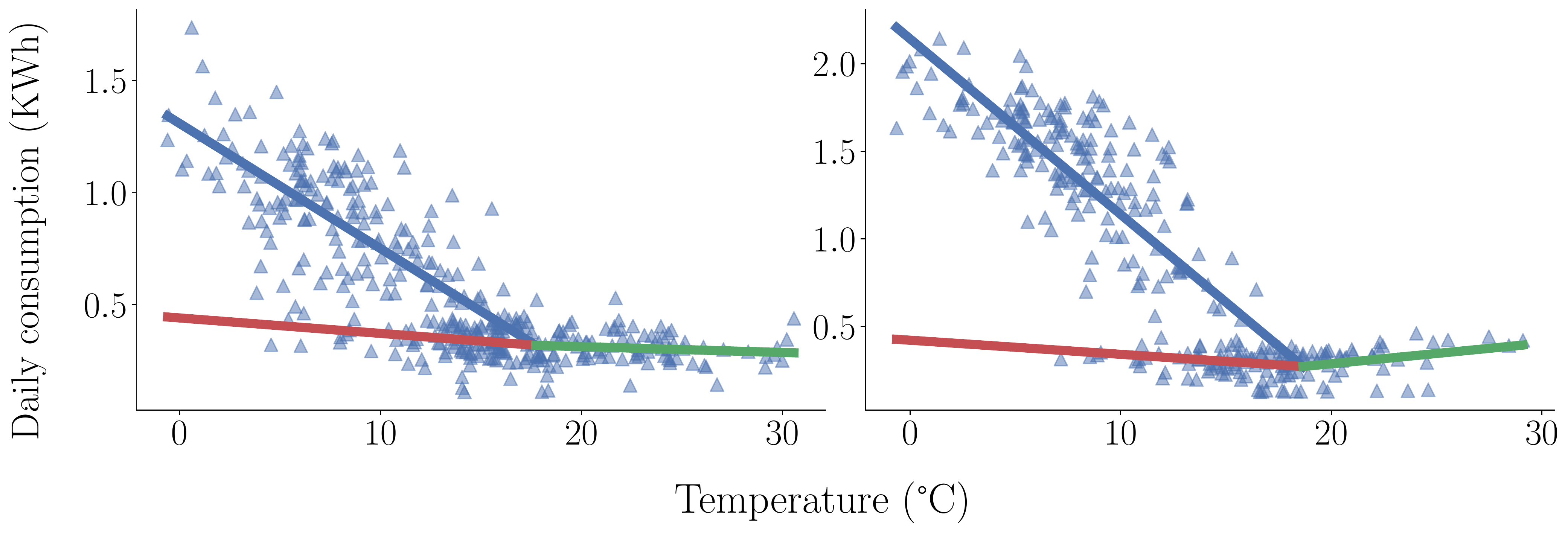}
  \caption[]{Example of households exhibiting a bi-modal consumption behaviour with two "home/away" states below a critical temperature. The data are fitted with the piece-wise mixture linear regression model.
  }
  \label{fig:pwml}
\end{figure}

\begin{figure}[!htb]
   \begin{minipage}{0.47\textwidth}
     \centering
     \includegraphics[width=1.0\linewidth]{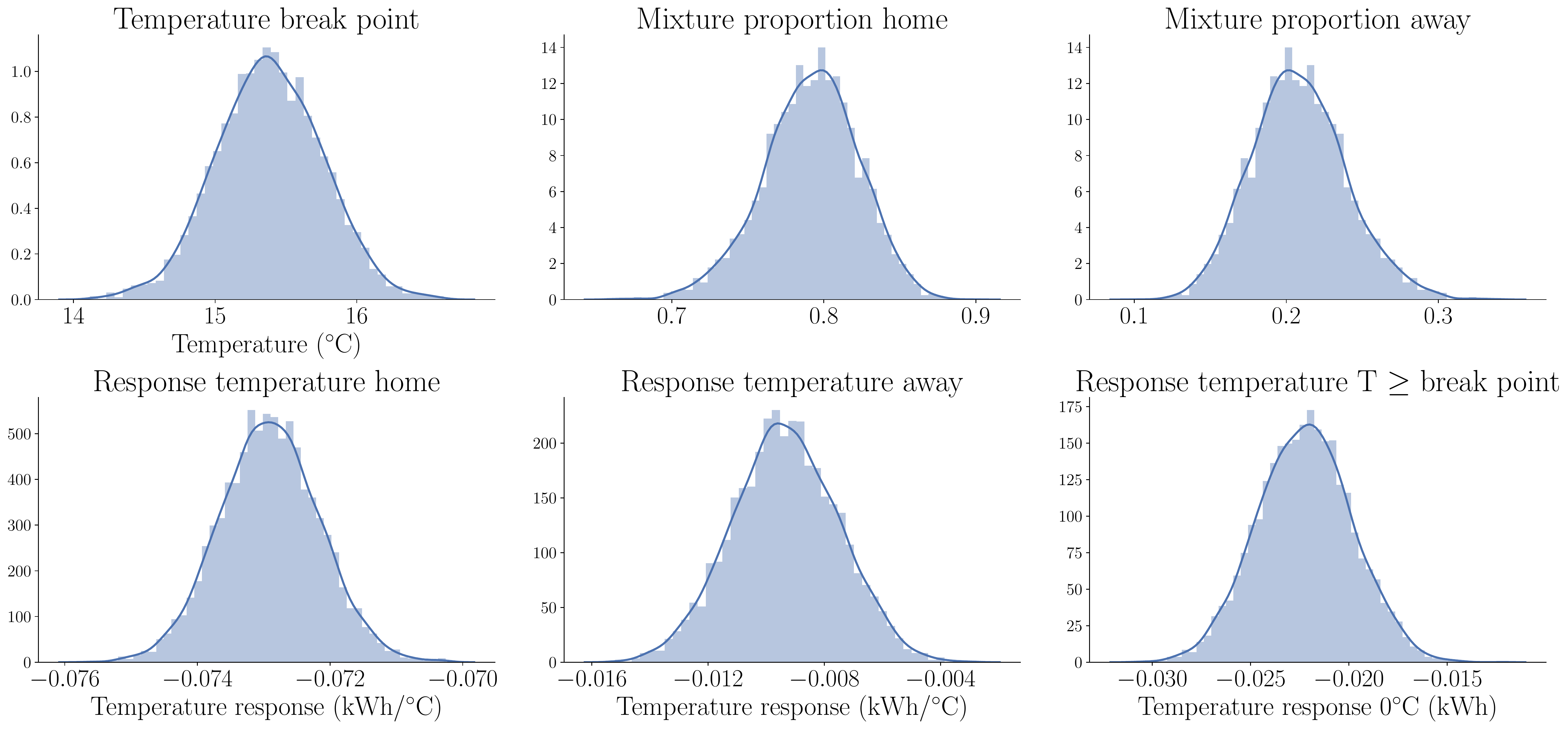}
     \caption{Posteriors obtained with the piece-wise mixture linear regression model for the critical temperature, mixture proportions for the "home/away" states and households responses to temperature for each hidden state.}\label{fig:posteriors}
   \end{minipage}
\end{figure}

Transformation $\tau$ is a combination of cumulative sum and inverse sigmoid transformations that is used to sample a random ordered tuple. We use the variational inference engine implemented in Pyro \cite{bingham2019pyro}. 
Latent state variable $z_i$ captures home/away states. A sketch of the graphical model is shown in \ref{fig:gm}. As an approximate posterior (guide) we use Normal distributions for all variables except for mixture weights, for which Dirichlet distribution is used.
In the current version observation scales of the three different branches are treated as parameters. We confirm the convergence by observing the loss and the parameters of the approximate posteriors. The posteriors and the best fits for exemplary households from the dataset are shown in Fig. \ref{fig:pwml}-\ref{fig:posteriors}.
We use the same prior parameters $w_{R, loc}$, $w_{R, scale}$, $b_{loc}$, $b_{scale}$, $\alpha_T$, $\alpha_s$, $\alpha_\omega$ across Hello Watt dataset. 

We note that the model described above is merely an example, and the formalism admits setting an arbitrary number of temperature threshold $T_k$ as well as an arbitrary number of components of mixture. This generalisation is potentially useful in case the consumption modes are different between cold, intermediate and hot regimes, or when we can identify more than two latent states, for example a family of two receiving two guests for a week. For the present usage and in the following we consider only one critical temperature threshold: $T_c$.

\section{Heating disaggregation}

For each household fitted with the mixture model, we estimate the consumption fraction due to heating $c^{(h)}$: for a given tuple $(c^{(tot)}, T)$ we evaluate the expectation $E\left[c^{(h)} \ | \ c^{(tot)}\right]$ and the variance $Var \left[c^{(h)} \ | \ c^{(tot)} \right]$ of consumption due to heating.

\begin{figure}
  \begin{subfigure}[b]{0.46\textwidth}
    \includegraphics[width=\textwidth]{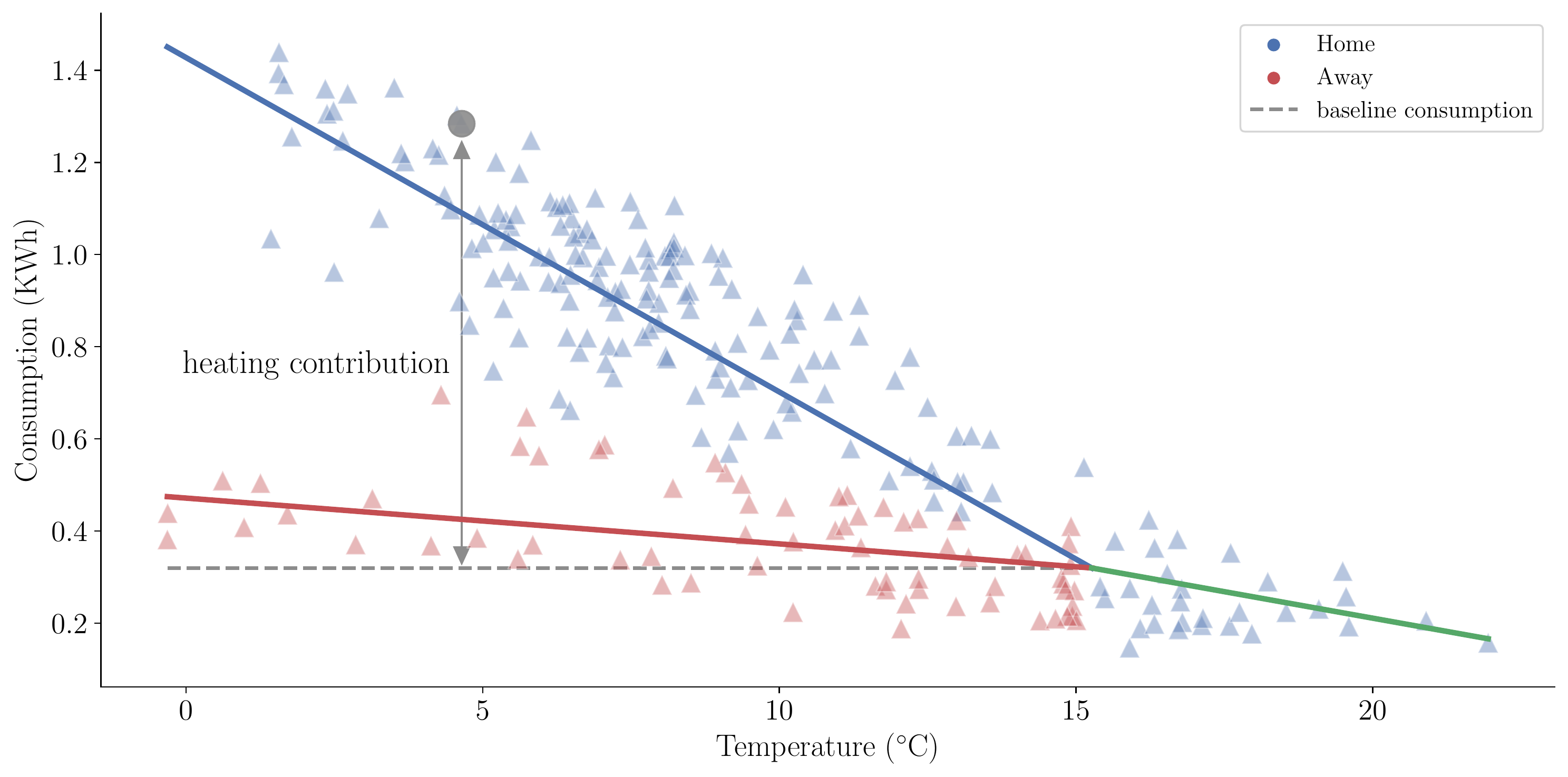}
    \label{fig:home_away_inferance}
  \end{subfigure}
  \centering
  \begin{subfigure}[b]{0.47\textwidth}
    \includegraphics[width=\textwidth]{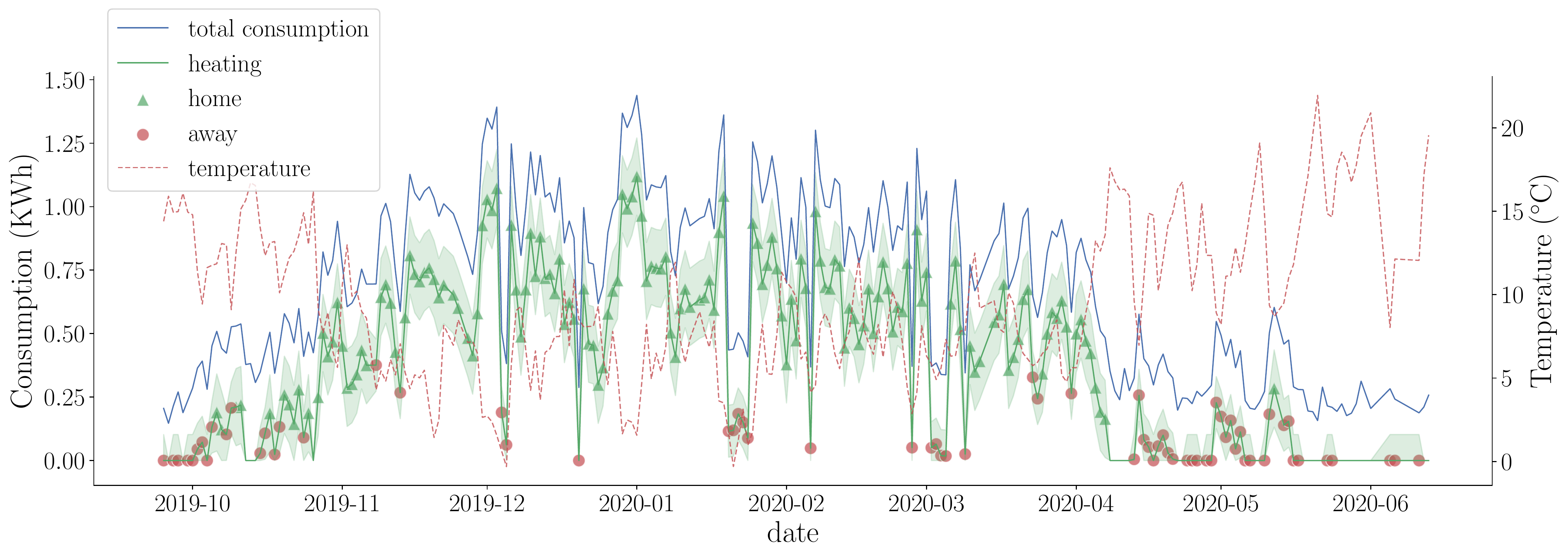}
    \label{fig:heating_disaggregation_home_away}
  \end{subfigure}
  \begin{subfigure}[b]{0.47\textwidth}
    \includegraphics[width=\textwidth]{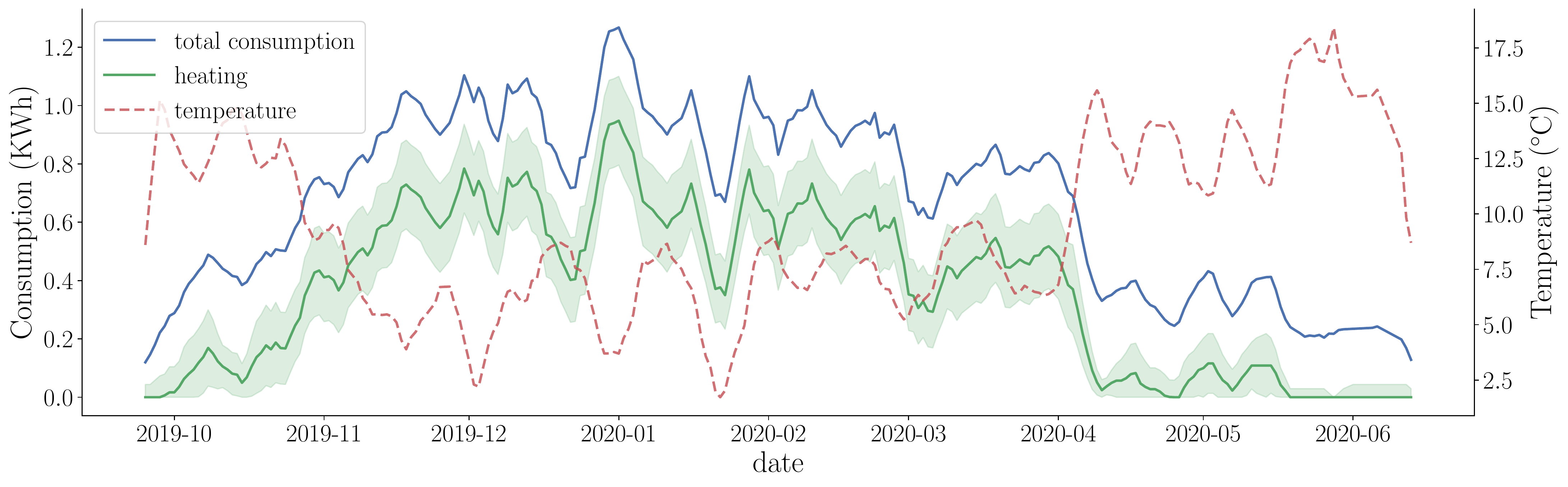}
    \label{fig:heating_disaggregation_moving_average}
  \end{subfigure}
  \caption{Example of heating disaggregation for one household. Top: temperature-consumption scatter plot, middle: consumption and temperature as a function of time, bottom: consumption and temperature as a function of time with a seven days moving average applied.}
  \label{fig:heating_disaggregation}
\end{figure}

We assume that the consumption below $T_c$ is a sum of the base $c(T_c) = w_a T_c + b_a$ \footnote{As $a$ we use the state for which bias $b$ is generated independently.} (assumed to be independent of current temperature and estimated at $T_c$) and of the heating $c^{(h)}$ parts, that together constitute $c^{(tot)}$ (cf. Fig. \ref{fig:heating_disaggregation}). We thus write:

$$c^{(h)} | \ c^{(tot)}, w_a, T_c,  b_a = c^{(tot)} - w_a T_c - b_a \ .$$

We then describe the distribution of $c^{(h)}$ using inferred posteriors $w_a$, $T_c$ and $b_a$, which are approximated by Normal distributions $w_a \sim \mathcal{N}(\mu_{w_a}, \sigma_{b_a})$, $T_c \sim \mathcal{N}(\mu_{T_c}, \sigma_{T_c})$, $b_z \sim \mathcal{N}(\mu_{b_z}, \sigma_{b_z})$, to obtain:
\begin{align*}
E\left[ c^{(h)} \right] &= c^{(tot)} - \mu_{w_a} \mu_{T_c} - \mu_{b_a} \ ,  \\ 
Var\left[ c^{(h)} \right] &= \left(\sigma^2_{w_a} + \mu^2_{w_a} \right) \left(\sigma^2_{T_c} + \mu^2_{T_c} \right) - \mu^2_{w_a}\mu^2_{T_c} + \sigma^2_{b_a} \ .
\end{align*}

Our approach also allows us to infer the latent state $z$ by maximizing the probability of the observed consumption to belong to the home/away state: 
$$ z^* = \underset{z\in \{home, away\}}{\operatorname{argmax}}
P(c^{(tot)} |  b_z, w_z, \sigma_z, T_c ) \ .$$

While neither the current temperature nor the home/away state enter the estimation of the heating fraction, the division of the consumption below $T_c$ into two states improves the overall quality of the fit and thus the inference of the heating consumption as well. An example of heating disaggregation is displayed in Fig. \ref{fig:heating_disaggregation}. 

To validate the heating disaggregation approach, we partition the consumption observation set for each household $j$ into two subsets $A$ and $B$, with observations before and after mid-January respectively. We consider as the ground truth the prediction for $c^{(h,AB)}$ for the heating part by fitting the model on $A\bigcup B$ and evaluating the heating contribution on $B$, and as the model prediction we consider the heating contribution $c^{(h,A)}$ estimated on $B$ with the model trained on $A$. To quantify the quality of the prediction we compute the relative root mean squared error (RMSE): 
$$\delta_j = \sqrt{\frac{1}{n_j}\sum \limits_p\left( \frac{c_{j, p}^{(h, AB)} - c_{j, p}^{(h, A)}} {c_{j, p}^{(h, AB)}}\right)^2} \ ,$$

where $n_j$ is the number of observations for the household $j$.
We report that the mean (over all the households) relative RMSE error $\delta_j$ is 16.6\% with the standard deviation of 13.9\%.

\section{Discussion and perspective}

In this paper we demonstrated a negative correlation between power consumption of households with electrical heating and external temperature and, based on this observation, constructed a Bayesian model for the disaggregation of the heating component.
Our Bayesian model allows to infer the response of the power consumption to external factors and identify modes of consumption in an unsupervised manner. 
The proposed model improves prior research\cite{Iyengar2018, Chambers2019} in the aspect a more detailed modeling: we introduce a mixture model that has the capacity to decode the occupancy state "home/away".

Thanks to the inference engine implemented in Pyro probabilistic programming language, our approach is also flexible with respect to the distribution choice, and so as a future step it would be natural to use LogNormal distributions instead of Normal distributions for the electrical consumption as prompted by our analysis of the energy consumption data. 

Our approach to graphical modeling of electrical consumption opens up potential applications: (a) description of properties of each housing category in a Bayesian way, (b) identification of outliers in each category (such as housings grouped by years of construction), (c) detection of latent electrical heating consumption (electrical heating housings not explicitly declared as such).

We also envision a more general approach to Bayesian modeling: rather than using one model to fit all possible instances of consumption data, we would specify a set of models, form mixtures from this set, and find the best fitting for each house, while penalising model complexity. 

From the point of view of improving the model by extending the number of features we intend to include wind speed and direction, and cloud coverage observations in the model, conditional on their significance.
The same mixture model can be applied to the daily gas consumption data from the smart meter Gazpar, for which data are currently being collected. From a practical point of view the knowledge of the two main heating energy sources will allow us to describe the heating consumption of the majority of the French households and identify housing insulation issues.

\bibliographystyle{ACM-Reference-Format}
\bibliography{bayesian_heating.bib}

\end{document}